\documentclass{aa} 
\usepackage{graphicx}
\usepackage{natbib}
\usepackage[english]{babel}
\usepackage[T1]{fontenc}
\usepackage[latin1]{inputenc}
\usepackage[dvips]{epsfig}

\begin{document}     

\title{CH$_3$OH and H$_2$O masers in high-mass star-forming regions}

\author{H. Beuther\inst{1} \and A. Walsh\inst{1} \and P. Schilke\inst{1} \and T.K. Sridharan\inst{3} \and K.M. Menten\inst{1} \and F. Wyrowski\inst{1}}

\institute{Max-Planck-Institut f\"ur Radioastronomie, Auf dem H\"ugel 69, 53121 Bonn, Germany \and Harvard-Smithsonian Center for Astrophysics, 60 Garden Street, MS 78, Cambridge, MA 02138, USA}

\offprints{H. Beuther,\\ \email{beuther@mpifr-bonn.mpg.de}}

\date{Received ... /Accepted ...}


\abstract{We present a comparison of Class {\sc ii} CH$_3$OH (6.7~GHz)
  and H$_2$O (22.2~GHz) masers at high spatial resolution in a sample
  of 29 massive star-forming regions. Absolute positions of both maser
  types are compared with mm dust continuum, cm continuum and
  mid-infrared sources. All maser features --~regardless of the
  species~-- are associated with massive mm cores, but only 3 out of
  18 CH$_3$OH masers and 6 out of 22~H$_2$O masers are associated with
  cm emission likely indicating the presence of a recently ignited
  massive star. These observations of a homogenous sample of massive,
  young star-forming regions confirm earlier results, obtained for each
  maser species separately, that both maser types are signposts of
  high-mass star formation in very early evolutionary stages. The data
  are consistent with models that explain CH$_3$OH maser emission by
  radiative pumping in moderately hot cores, requiring the absence, or
  only weak, free-free cm continuum radiation due to recently ignited
  stars. Mid-infrared sources are associated with both maser types in
  approximately $60\%$ of the observed fields. Thus, mid-infrared
  objects may power maser sites, but the detection of strong
  mid-infrared emission is not strictly necessary because it might be
  heavily extincted. A comparison of the spatial separations between
  the different observed quantities and other properties of the
  star-forming regions does not reveal any correlation. Our data
  suggest that CH$_3$OH and H$_2$O masers need a similar environment
  (dense and warm molecular gas), but that, due to the different
  excitation processes (radiative pumping for CH$_3$OH and collisional
  pumping for H$_2$O), no spatial correlations exist. Spatial
  associations are probably coincidences due to insufficient angular
  resolution and projection effects. The kinematic structures we find
  in the different maser species show no recognizable pattern, and we
  cannot draw firm conclusions whether the features are produced in
  disks, outflows or expanding shock waves. }

\maketitle


\section{Introduction}

Over the last decade, it has become clear that Class {\sc ii} CH$_3$OH
and H$_2$O maser emission is closely associated with the earliest
stages of massive star formation. Much research has been focused on
both maser species, but most high-resolution studies have concentrated
on only one type. Little is known about the associations and
connections of the CH$_3$OH and the H$_2$O maser emission. Motivated
by this lack of information, we conducted high-resolution observations
of both maser types in a large sample of high-mass protostellar
objects (HMPOs) and discuss them in the context of different other
probes of massive star formation.

So far, high-resolution investigations, focused either on H$_2$O or on
CH$_3$OH maser emission, indicate independently that both
maser types are found in outflows as well as in circumstellar
disks.  Examples of these studies on the H$_2$O side are, e.g.,
\citet{torrelles 1997,torrelles 1998}, \citet{codella 1995}, and 
\citet{codella 1996,codella 1997}, whereas recent studies of CH$_3$OH 
masers include \citet{walsh 1997,walsh 1998}, \citet{phillips 1998},
\citet{minier 2000,minier 2001}, and \citet{codella 2000}. Reviews on
this subject are found in \citet{menten 1996}, \citet{garay 1999},
\citet{norris 2000}, and \citet{kylafis 1999}.

Another finding is that both maser types are occasionally spatially
coincident with ultracompact H{\sc ii} regions, but more often with
hot molecular cores and/or massive molecular outflows \citep{hofner
1996,codella 1996,codella 1997,walsh 1998,garay 1999,minier
2001}. Recent studies by \citet{walsh 2001} indicate that CH$_3$OH
masers often are associated with mid-infrared sources that should be
luminous enough to produce strong ionizing radiation. In spite of
that, a number of those maser sites is undetected in the radio
continuum, and \citet{walsh 2001} conclude that these might be in an
evolutionary stage prior to forming an ultracompact H{\sc ii} region.

For sources in which both maser species are observed (mostly
single-dish data with low spatial resolution), the velocity spread is
far higher in the case of H$_2$O maser emission than known for
CH$_3$OH masers. For example, \citet{sridha} found in a sample of
high-mass protostellar candidates that the CH$_3$OH emission is
confined always to a total radial velocity extent of at most 15 km/s,
whereas the H$_2$O maser emission is found extending up to 70 km/s. In
spite of the similarities outlined above, these different spectral
features show that the two maser types are subject to different
excitation conditions and likely occur in separate spatial regions
around the evolving massive protostars.

The current status of CH$_3$OH maser excitation studies favors a
scenario in which the CH$_3$OH maser emission is produced by radiative
pumping in small, moderately hot regions ($\sim 150$~K) with CH$_3$OH
column densities $>10^{15}$~cm$^{-2}$ and hydrogen number densities
$<10^8$~cm$^{-3}$ \citep{hartquist 1995,sobolev 1997,cragg
2001}. While it was suggested that CH$_3$OH maser could be produced in
disks (e.g., \citealt{norris 1998}), this is not unambiguously
confirmed, and other scenarios like expanding shock waves or outflows
have been proposed as well (e.g., \citealt{walsh 1998},
\citealt{minier 2000b}). It is quite possible that the same maser 
species in different sources have different origins.

In contrast, most H$_2$O maser models explain the excitation by
collisional pumping with H$_2$ molecules within shocks associated with
outflows and/or accretion \citep{elitzur 1989, garay 1999}. For
example, the model by \citet{elitzur 1989} requires dissociative
shocks in dense gas (preshock densities around $10^7$~cm$^{-3}$) with
temperatures up to $\sim 400$~K. In many sources, it is believed that
the H$_2$O masers form within outflows, but there are also examples
claiming that the H$_2$O masers are produced in disks \citep{garay
1999}.

The high-resolution study of H$_2$O and CH$_3$OH maser emission in 29
high-mass star-forming regions presented here intends to compare the
characteristics of these important signposts of massive star
formation. The study is part of a long term project studying very
early evolutionary stages of massive star formation, prior to, or
during the formation of an ultracompact H{\sc ii} region. The entire
sample of 69 sources presented first by \citet{sridha} is distributed
mainly between 1~kpc and 10~kpc distance from the sun. For many
sources, only kinematic distances are known, and the data suffer from
the distance ambiguity of this approach
\citep{sridha}. The bolometric luminosities of the sources range
roughly between $10^{3.5}$~L$_{\odot}$ and $10^{5.6}$~L$_{\odot}$, and
the core masses vary between a few 100~M$_{\odot}$ and a few
$10^4$~M$_{\odot}$. Detailed analyzes of different sample properties
(luminosities, density distributions, outflows) are presented in
\citet{sridha} and \citet{beuther 2001a,beuther 2001b}. In spite of the
non-detection in Galaxy-wide single-dish cm surveys, more sensitive
interferometric studies revealed that in some of the sources cm
emission occurs, whereas in others we do not find any free-free
emission down to 1~mJy. This makes the sample ideally suited for the
study of associations of masers with different kinds of sources and
evolutionary stages. The basic source properties of the sub-sample
analyzed in this paper are summarized in Table \ref{sample}. We
compare the maser emission mainly with mm dust emission tracing the
massive core \citep{beuther 2001a}, cm wavelength emission
believed to be due to free-free emission in those sources where it had
been detected \citep{sridha}, and mid-infrared emission likely due to
warm dust around embedded objects.
 

\section{Observations}

\subsection{Single-dish H$_2$O and CH$_3$OH maser data}

An H$_2$O (22.2~GHz) and Class {\sc ii} CH$_3$OH Class (6.7~GHz) maser
survey of the the whole sample of 69 high-mass protostellar objects,
as introduced by \citet{sridha}, was conducted with the Effelsberg
100~m telescope in two runs in 1998. Altogether, 29 H$_2$O and 26
CH$_3$OH masers were found. The observations are described and
statistically analyzed in \citet{sridha}. Follow-up observations of
the maser detections were conducted with high spatial resolution with
the VLA and the ATCA.

\subsection{H$_2$O maser observations with the VLA}

High-resolution images of the 22.2~GHz H$_2$O maser line were obtained
with the Very Large Array (VLA) in the B configuration with a spectral
resolution of 0.65~km/s, a synthesized HPBW of $\approx 0.4''$, and a
primary beam FWHM of $\approx 2'$. The snapshot mode was used with $3
\times 5$ minutes integrations spread over the transit of each
source. We reduced the data with the AIPS software package using
standard procedures. The typical $1\sigma$ sensitivity is 1~Jy.

For 22 out of 26 observed sources, the signal-to-noise ratio and the UV
coverage were sufficiently good to derive absolute positions to an
accuracy better than $1''$. The accuracy of relative positions of
different maser features is around $0.1''$.

\subsection{CH$_3$OH maser observations with the ATCA}

To get high-resolution images of the Class {\sc ii} 6.7~GHz maser line
of CH$_3$OH, we employed the Australian Telescope Compact Array (ATCA)
which is part of the Australia Telescope National Facility. From the
26 sources initially found to be associated with CH$_3$OH maser
emission, 21 are south of declination 20$^{\circ}$ which is necessary
to be observable with the ATCA. Six of the 21 sources had already been
observed with the ATCA by \citet{walsh 1998}. Thus, we observed the
remaining 15 objects in the snapshot mode with a series of 6 short
integration cuts of 5 minutes for each source. Phase calibrators were
observed every 20 minutes. The field of view of the ATCA at this
frequency is $\sim(8')^2$, and the effective HPBW of the synthesized
beam is $\sim 1.9''$. Observations near the equator result in
elongated, elliptical beams because of the east--west elongation of
the array. The spectral resolution was $\sim 0.2$~km/s, and the
$1\sigma$ sensitivity was typically 0.5~Jy.

In 10 out of the 15 sources the data are sufficiently good to derive
positions. The data reduction was performed with the MIRIAD software
package using standard procedures. Absolute positions are estimated to
be accurate to 1$''$ whereas the relative positions should be accurate
to $\sim 0.1''$. In 4 sources the signal was too weak to be detected,
whereas in 1 source the signal was strong enough, but so close to the
equator that no positions could be derived.

\subsection{Other observations}

{\it Dust emission at mm wavelengths}

The 29 sources presented in this paper were observed as part of the
large sample of 69 high-mass protostellar objects introduced by
\citet{sridha} with the MAMBO array in the 1.2~mm continuum. A
detailed description of the data reduction and analysis can be found
in \citet{beuther 2001b}. For a few sources even higher resolution mm
data were obtained with the Plateau de Bure Interferometer and the
BIMA array at 2.6~mm. Plateau de Bure data are used here for
05358+3543 and 19217+1651, whereas the BIMA data were used in the
cases of $18089-1733$, $18264-1152$, 18566+0408 and 23033+5951.
Details of the Plateau de Bure observations are presented in
\citet{beuther 2001c} and Beuther et al. (in prep.), and the BIMA
observations will be published by Wyrowski et al. (in prep.).

The pointing accuracy of the MAMBO array is estimated to be around $5''$
and the BIMA and Plateau de Bure data should be accurate to
approximately $1''$.\\

\noindent {\it Continuum emission at cm wavelengths, likely produced by
free-free emission}

To know whether faint cm emission is present in some of the sources,
we obtained 3.6~cm continuum images with the VLA down to a $1\sigma$
sensitivity of 0.1~mJy with a spatial resolution of $0.7''$. For more
details see \citet{sridha}. The positions of these data are correct to
within $1''$.\\

\noindent {\it Mid-infrared data}

The Midcourse Space Experiment (MSX) point source catalog was searched
for mid-infrared sources in the respective fields of view. For details
on MSX see, e.g., \citet{egan 1998} and the MSX Web
site\footnote{http://www.ipac.caltech.edu/ipac/msx/msx.html}. In most
fields, mid-infrared sources were found \citep{sridha}. For the
few sources, for which the point source catalog had no entries, we
examined the most sensitive MSX images at 8~$\mu$m and extracted source
positions. The spatial resolution of MSX is approximately $18''$, and
the sensitivity varies between the different bands between 0.1 and
30~Jy/beam. \citet{egan 1998} quote the positions to be accurate
within $4''-5''$. For 05358+3543, we use a mid-infrared source
position observed by McCaughrean et al. (in prep.) with the Keck
observatory at 11.7~$\mu$m.


\section{Results}
\label{overall}

\subsection{The database}
\label{database}

Figure \ref{maser1} presents maps of the 29 maser sources
observed with high spatial resolution. For 6 sources the CH$_3$OH data
are taken from \citet{walsh 1998}, for 19410+2336 the CH$_3$OH
maser positions are presented by \citet{minier 2000}, and for
05358+3543 no high-resolution CH$_3$OH maser position is known (single
dish detection by, e.g., \citealt{sridha} and \citealt{menten
1991}). Of the 6 sources in our sample north of 20$^\circ$
declination, Effelsberg 100~m data show that three (19410+2336,
20126+4104 \& 23139+5939) contain CH$_3$OH masers (Table
\ref{sample}), but because the ATCA is only capable of observing
sources below 20$^\circ$ declination, we have accurate CH$_3$OH maser
positions only for 19410+2336 and 20126+4104, both of which were
observed with the European VLBI network (EVN) by \citet{minier
2000,minier 2001}. Table \ref{positions} lists the positions of the
maser features.

As we want to understand what kind of source is powering the masers,
Fig \ref{maser1} also shows other features characterizing the massive
star-forming regions:\\ (1) mm dust continuum data, tracing the
massive molecular cores in which the stars form,\\ (2) mid-infrared
data, indicating warm dust around an embedded protostar,\\ (3) and cm
continuum emission, probably due to optically thin free-free emission
signposting recently ignited massive stars belonging to the same
cluster. But as we do not have spectral index information in the cm
regime, it is also possible that some of the cm features are still
optically thick or could be due to jets (see also \citealt{sridha}).

Table \ref{sample} presents more characteristic source properties,
such as the total luminosities, distances and core masses
\citep{sridha,beuther 2001a}.

\begin{figure*}
\caption{CH$_3$OH (squares) and H$_2$O (triangles) maser positions are
 presented. The grey-scales with contours show the mm continuum images
 (thermal dust emission), the black (on white) contours present the cm
 observations tracing the free-free emission and the stars mark the
 mid-infrared point source positions. The axes show offsets in
 arcseconds from the IRAS position with varying scales for the
 different sources. \label{maser1}}
\end{figure*}

\setcounter{figure}{0}

\begin{figure*}
\caption{continued}
\end{figure*}

\setcounter{figure}{0}

\begin{figure}
\caption{continued}
\end{figure}
 
\begin{table*}[ht]
\begin{center}
\begin{tabular}{lrrrrrrrrrr}
source & $d_{\rm{far}}$ & $d_{\rm{near}}$ & $L_{\rm{far}}$ & $L_{\rm{near}}$ & $M_{\rm{far}}$ & $M_{\rm{near}}$ & $N_{\rm{H_2}}$ & cm & CH$_3$OH & H$_2$O \\
 & [kpc] & [kpc] & [L$_{\odot}$] & [L$_{\odot}$] & [M$_{\odot}$] & [M$_{\odot}$] & [10$^{23}$cm$^{-2}$] & [mJy] & [Jy] & [Jy] \\
\hline
 05358+3543  &   1.8  &    --    &  3.8   &   --    &   610     &    --     &   5.8   &     $<1$   &      162   &     45     \\
 18089$-$1732  &   13.0 &   3.6    &  5.6   &   4.5   &   31900   &    2400   & 17    &     0.9    &      54    &     75   \\
 18090$-$1832  &   10.0 &   6.6    &  4.5   &   4.1   &   4600    &    2000   & 2.4   &     ?      &      82    &     --   \\
 18102$-$1800  &   14.0 &   2.6    &  5.3   &   3.8   &   24000   &    800    & 2.9   &     44     &       9    &     --   \\
 18151$-$1208  &   3.0  &   --     &  4.3   &   --    &   1100    &    --     & 4.4   &     $<1$   &      50    &    0.8   \\
 18182$-$1433  &   11.8 &   4.5    &  5.1   &   4.3   &   20700   &    3000   & 9.4   &     $<1$   &      24    &      18  \\
 18264$-$1152  &   12.5 &   3.5    &  5.1   &   4.0   &   55300   &    4300   & 16    &     $<1$   &      4     &      50  \\
 18290$-$0924  &   10.5 &   5.3    &  5.0   &   4.4   &   6500    &    1600   & 3.3   &     7.0    &      15    &      4   \\
 18306$-$0835  &   10.7 &   4.9    &  4.8   &   4.1   &   16400   &    3400   & 6.9   &     82     &      --    &    0.7   \\
 18308$-$0841  &   10.7 &   4.9    &  4.9   &   4.2   &   12800   &    2700   & 4.6   &     $<1$   &      --    &     1.5  \\
 18310$-$0825  &   10.4 &   5.2    &  4.8   &   4.2   &   15300   &    3800   & 1.8   &     7.0    &      --    &    --    \\
 18345$-$0641  &   9.5  &    --    &  4.6   &   --    &   6900    &    --     & 2.4   &     27     &      5     &     3    \\
 18372$-$0541  &   13.4 &   1.8    &  5.3   &   3.5   &   10200   &    200    & 2.6   &     80     &      9     &    1.5   \\
 18385$-$0512  &   13.1 &   2.0    &  5.3   &   3.7   &   13400   &    300    & 4.5   &     29     &      --    &     200  \\
 18440$-$0148  &   8.3  &    --    &  4.7   &   --    &   1700    &    --     & 0.4   &     $<1$   &      4     &      ?   \\
 18488+0000  &   8.9  &   5.4    &  4.9   &   4.5   &   6100    &    2200   &   2.9   &     194    &      17    &     1      \\
 18521+0134  &   9.0  &   5.0    &  4.6   &   4.1   &   3000    &    900    &   2.0   &     $<1$   &       1    &     --     \\
 18553+0414  &   12.9 &   0.6    &  5.1   &   2.4   &   15800   &    35     &   3.6   &      $<1$  &      --    &    50      \\
 18566+0408  &   6.7  &   --     &  4.8   &   --    &   2100    &    --     &   2.9   &     $<1$   &      7     &     3      \\
 19012+0536  &   8.6  &   4.6    &  4.7   &   4.2   &   3900    &    1100   &   2.9   &     $<1$   &      1     &     2      \\
 19035+0641  &   2.2  &    --    &  3.9   &   --    &   390     &    --     &   1.9   &     4.0    &       14   &     9      \\
 19217+1651  &   10.5 &    --    &  4.9   &   --    &   9500    &    --     &   5.3   &      32    &       1    &     9      \\
 19282+1814  &   8.2  &   1.9    &  4.9   &   3.6   &   2700    &    150    &   2.4   &     $<1$   &       3    &     --     \\
 19410+2336  &   6.4  &   2.1    &  5.0   &   4.0   &   7800    &    800    &   5.7   &     1      &       30   &     110    \\
 20126+4104  &   1.7  &    --    &  3.9   &   --    &   500     &    --     &   5.2   &     $<1$   &       36   &     15     \\
 20293+3952  &   2.0  &   1.3    &  3.8   &   3.4   &   500     &    200    &   1.9   &     7.6    &       --   &     100    \\
 23033+5951  &   3.5  &    --    &  4.0   &   --    &   2300    &    --     &   3.7   &     1.7    &       --   &     4      \\
 23139+5939  &   4.8  &    --    &  4.4   &   --    &   1800    &    --     &   4.0   &     1.4    &       3    &     400    \\
 23151+5912  &   5.7  &    --    &  5.0   &   --    &   1200    &    --     &   1.8   &     $<1$   &        --  &     60     \\ 
\hline
\end{tabular}
\end{center}
\caption{Characteristic parameters of the maser associated 
massive star-forming regions. The distances, luminosities, 3.6~cm
fluxes and single-dish maser fluxes are taken from \citet{sridha}, the
core masses and peak column densities of the main mm core are derived
from 1.2~mm dust continuum data \citep{beuther 2001a}. If no
near-values are given, the distance ambiguity is solved
\citep{sridha}. \label{sample}}
\end{table*}

\begin{table*}[ht]
\begin{tabular}{lrrrr}
source & CH$_3$OH & $v_{\rm{LSR}}$ & H$_2$O & $v_{\rm{LSR}}$  \\
       & [J2000.0]& [km/s]  & [J2000.0]& [km/s]  \\
\hline
05358+3543 & \citet{menten 1991},     &       & 05:39:13.0 35:45:48.7  & $-$17      \\
           & single-dish obs.         &       &                       &          \\
18089$-$1732 & 18:11:51.4 $-$17:31:30.2$^1$& 39   & 18:11:51.5 $-$17:31:28.8& 32       \\
18090$-$1832 & 18:12:01.9 $-$18:31:55.0$^1$& 108  &                         &          \\
18102$-$1800 & 18:13:11.3 $-$17:59:57.8 & 25    &                         &          \\
      & 18:13:11.3 $-$17:59:58.0 & 24/23 &                         &          \\
18151$-$1208 & 18:17:58.1 $-$12:07:25.4 & 28    & 18:17:50.3 $-$12:07:52.9   & 22/25/32 \\
18182$-$1433 & 18:21:09.2 $-$14:31:48.6$^1$& 62   & 18:21:09.1 $-$14:31:48.7 & 60       \\
           &                          &       & 18:21:09.1 $-$14:31:47.8 & 65       \\
           &                          &       & 18:21:09.0 $-$14:31:48.5 & 57       \\
           &                          &       & 18:21:09.0 $-$14:31:47.9 & 50/53    \\
           &                          &       & 18:21:09.1 $-$14:31:48.6 & 68       \\
18264$-$1152 & 18:29:14.4 $-$11:50:22.4 & 47    & 18:29:14.2 $-$11:50:22.6  & 45/52    \\
           &                          &       & 18:29:14.4 $-$11:50:24.5& 61/68/74 \\
           &                          &       & 18:29:14.3 $-$11:50:22.5& 52       \\
18290$-$0924 & 18:31:44.2 $-$09:22:12.5$^1$& 80   & 18:31:44.1 $-$09:22:12.1& 88       \\
18306$-$0835 &                          &       & 18:33:23.9 $-$08:33:31.6& 105      \\
           &                          &       & 18:33:17.1 $-$08:33:24.3& 82       \\
18308$-$0841 &                          &       & 18:33:33.2 $-$08:39:15.1& 69/73/86 \\
18310$-$0825 & 18:33:43.8 $-$08:21:20.5$^1$& 88  &                         &          \\
18345$-$0641 & 18:37:16.9 $-$06:38:30.4  & 97    &                         &          \\
           & 18:37:17.0 $-$06:38:30.3  & 94    &                         &          \\
18372$-$0541 & 18:39:55.9 $-$05:38:45.1  & 25    & 18:39:55.9 $-$05:38:44.1  & 23       \\
           & 18:39:55.9 $-$05:38:44.9  & 24    & 18:39:55.9 $-$05:38:44.6  & 26       \\
           & 18:39:56.0 $-$05:38:45.9  & 23    &                         &          \\
           & 18:39:56.0 $-$05:38:45.8  & 18    &                         &          \\
18385$-$0521 &                          &       & 18:41:13.2 $-$05:09:00.5 & 19       \\
           &                          &       & 18:41:13.2 $-$05:09:01.0 & 11       \\
           &                          &       & 18:41:13.2 $-$05:09:01.1 &  26      \\
           &                          &       & 18:41:13.2 $-$05:09:01.2 & 30       \\
           &                          &       & 18:41:13.2 $-$05:09:01.3 & 37       \\
18440$-$0148 & 18:46:36.7 $-$01:45:22.2$^1$& 105  &                         &          \\
18488+0000 &                          &       & 18:51:24.5 00:04:33.7   & 80       \\
18521+0134 & 18:54:40.8 01:38:04.5   & 77    &                         &          \\
18553+0414 &                        &       & 18:57:53.4 04:18:17.4   & 4        \\
           &                        &       & 18:57:53.4 04:18:17.3   & 12       \\
18566+0408 & 18:59:10.0 04:12:14.7   & 87/86 & 18:59:10.0 04:12:15.6   & 88       \\
           & 18:59:10.0 04:12:14.1   & 84    & 18:59:10.0 04:12:15.7   & 68       \\
           & 18:59:10.0 04:12:14.0   & 79    &                         &          \\
\hline
\end{tabular}
\end{table*}
\begin{table*}[h]
\begin{tabular}{lrrrr}
source & CH$_3$OH & $v_{\rm{LSR}}$ & H$_2$O & $v_{\rm{LSR}}$    \\
       & [J2000.0]& [km/s]  & [J2000.0]& [km/s]  \\
\hline
19012+0536 &                        &       & 19:03:45.3 05:40:42.8& 63          \\
19035+0641 & 19:06:01.6 06:46:35.9   & 37/31 & 19:06:01.6 06:46:36.3& 21          \\
19217+1651 & 19:23:58.9 16:57:41.8  & 7     & 19:23:58.5 16:57:41.4& 18          \\
           &                        &       & 19:23:58.8 16:57:41.5& 11          \\
           &                        &       & 19:23:58.8 16:57:41.1& 8           \\
19282+1814 & 19:30:23.0 18:20:27.1  & 19    &                      &             \\
19410+2336 & 19:43:11.3 23:44:03.3$^2$& 27  & 19:43:11.2 23:44:03.0& 27          \\
           & 19:43:11.2 23:44:03.0$^2$& 17  & 19:43:11.2 23:44:03.1& 19          \\
20126+4104 & 20:14:26.0 41:13:33.4$^2$& $-$6  & 20:14:26.0 41:13:32.6& 0           \\
           &                        &       & 20:14:26.0 41:13:32.7& $-$8          \\
20293+3952 &                        &       & 20:31:12.9 40:03:22.8& 2/$-$28       \\
23033+5951 &                        &       & 23:05:25.0 60:08:14.1& $-$49/$-$55/$-$62 \\
23139+5939 &                        &       & 23:16:10.3 59:55:28.7& $-$48/$-$53     \\
           &                        &       & 23:16:10.3 59:55:28.6& $-$44         \\
           &                        &       & 23:16:10.4 59:55:28.8& $-$33         \\
23151+5912 &                        &       & 23:17:20.8 59:28:47.0& $-$52         \\
           &                        &       & 23:17:20.3 59:28:47.3& $-$69         \\
\hline      
\end{tabular}
\caption{Positions of the CH$_3$OH \& H$_2$O maser. For the sources with citation we took the CH$_3$OH maser data from $^1$\citet{walsh 1998} and $^2$\citet{minier 2000,minier 2001}.}
\label{positions}
\end{table*}

It is of major interest to see how the different maser types are
associated with other observable features. Because of the positional
accuracy of the maser and cm observations of $\approx 1''$, we cannot
reliably separate features which are closer than $1.5''$ to each
other. On linear scales at the given distances of a few kpc, this is
still a large separation (e.g., at 4~kpc distance, $1.5''$ correspond
to $\approx 6000$~AU~$\approx 0.03$~pc), and we cannot be certain
whether sources within $1.5''$ of each other are associated or not. It
has also to be taken into account that those separations are
projections on the plane of the sky, and there are probably spatial
offsets along the line of sight as well. If more than one maser
feature is detected we derived the separations using the nearest
positions (except of the H$_2$O maser spots in 19217+1651 because
clearly two different groups exist). \citet{lada 1999} report a mean
separation between stellar objects in clusters around 0.1~pc, and
\citet{mccaughrean 1994} find mean separations in the center of the 
Orion Nebula Cluster of $\sim0.03$~pc. The scenario of merging
intermediate-mass protostars within the innermost center of
star-forming cores requires even higher protostellar densities between
$10^6$~pc$^{-3}$ and $10^8$~pc$^{-3}$, corresponding to mean
separations between 0.01~pc and 0.002~pc \citep{bonnell 1998,stahler
2000}. Taking all this different estimates into account, it is
possible that a significant fraction of features, which seem to be
associated within our observational accuracy, are powered by the same
underlying source, but we cannot be certain, and it is possible that
multiple features are powered by separate multiple sources. Within
these powering sources, different maser origins are possible, e.g.,
outflows, disks or expanding shock waves (e.g.,
\citealt{garay 1999}). Furthermore, separate components of a binary or
multiple system might account for the maser emission, because multiple
system are more common in high-mass than in low-mass star-forming
regions \citep{preibisch 1999}. Due to the lower spatial resolution and lower
positional accuracy, mm and mid-infrared sources cannot be separated
from the maser features when they are located within $5''$.

As the sample is distributed throughout the whole Galactic disk with
largely different distances, we calculate linear separations between
the masers and the other observable quantities (mm, cm and
mid-infrared emission). For sources, where the separation is
below the pointing accuracy of the observations, the derived values
have to be taken cautiously because the errors are larger.
As we only know kinematic distances for many of the sources,
Tables \ref{ch3oh_sep} and \ref{h2o_sep} list the separations
derived for the near and the far distance, respectively. Table
\ref{association} presents the mean separations; for sources with distance 
ambiguity we choose the near distance for this
calculation. Additionally, a summary of the features which are,
within the observing accuracy, associated with each other is given
in Table \ref{association}.

\begin{table*}[ht]
\begin{tabular}{lrrrrrrrrrrrr}
source & \multicolumn{3}{c}{CH$_3$OH/H$_2$O} & \multicolumn{3}{c}{CH$_3$OH/mm} & \multicolumn{3}{c}{CH$_3$OH/cm} & \multicolumn{3}{c}{CH$_3$OH/MIR} \\
&      & far  & near  &     & far  & near  &     & far  & near  &     & far  & near \\
& [$''$] & [pc] & [pc] & [$''$] & [pc] & [pc] & [$''$] & [pc] & [pc] & [$''$] & [pc] & [pc] \\ 
\hline
18089$-$1732  & 1.5 & 0.1 & 0.03 & 1.5 & 0.1 & 0.03 & 1.5 & 0.1 & 0.03 & 11 & 0.69 & 0.19 \\
18090$-$1832  &     & & & 0.5  & 0.02 & 0.02 &     & & & 0.5 & 0.02 & 0.02 \\
18102$-$1800  &     & & & 4   & 0.27 & 0.05 & 30  & 2.04 & 0.38 & 30 & 2.04 & 0.38 \\
18151$-$1208  & 80  & 1.16 & & 0.5  & 0.01 & &     & & & 12 & 0.16 & \\
18182$-$1433  & 1   & 0.06 & 0.02 & 1   & 0.06 & 0.02 &     & & & 12 & 0.67 & 0.26\\     
18264$-$1152  & 1   & 0.06 & 0.02 & 1   & 0.06 & 0.02 &     & & & 9  & 0.55 & 0.15\\
18290$-$0924  & 1   & 0.05 & 0.03 & 2   & 0.1 & 0.05 & 20  & 1.02 & 0.51 & 1  &0.05 & 0.03\\
18310$-$0825  &     & & & 4   & 0.2 & 0.1 &     & & & 16 & 0.81 & 0.40 \\
18345$-$0641  &     & & & 0.5 & 0.02& &     & & & 3  & 0.14 & \\
18372$-$0541  & 0.5  & 0.03 & 0.004 & 1   & 0.07& 0.01 & 1   & 0.07 & 0.01 & 1  &0.07 &0.01 \\
18440$-$0148  &     & & & 1   & 0.04 & &     & & & 4  & 0.16 & \\    
18521+0134    &     & & & 1   & 0.04 & 0.02 &    & & & 4  & 0.18 & 0.1 \\
18566+0408    & 1   & 0.03 & & 1   & 0.03 & &     & & & 1  & 0.03 & \\
19035+0641    & 0.5  & 0.01 & & 2   & 0.02 & & 7   & 0.08 & & 1  & 0.01 & \\
19217+1651    & 0.5 & 0.03 & & 1   &0.05 & & 6   &0.3 & &1.5 &0.08 & \\
19282+1814    &     & & & 1   & 0.04 & 0.01 &     & & & 20 & 0.80 & 0.18 \\
19410+2336    & 0.5  & 0.02 & 0.01 & 5   & 0.16 & 0.05 & 1   & 0.03 & 0.01 & 5  & 0.16 & 0.05\\
20126+4104    & 1   & 0.01 & &  1  & 0.01 & &     & & &  3 & 0.03 & \\
\hline
\end{tabular}
\caption{Separations of the CH$_3$OH masers from H$_2$O, mm, cm and 
mid-infrared sources. Far and near values correspond to the far and
near distance, respectively. For sources without near values, the
distance ambiguity is solved \citep{sridha}. The {\bf uncertainty} of
the CH$_3$OH/H$_2$O column is $\sim 1.5''$ and scales with the
distances for the linear separations. The {\bf uncertainties} of the
other columns are $\sim 5''$, scaling with the distances as
well. \label{ch3oh_sep}}
\end{table*}

\begin{table*}[ht]
\begin{tabular}{lrrrrrrrrr}
source & \multicolumn{3}{c}{H$_2$O/mm} & \multicolumn{3}{c}{H$_2$O/cm}
& \multicolumn{3}{c}{H$_2$O/MIR} \\ & & far & near & & far & near & &
far & near \\ & [$''$] & [pc] & [pc] & [$''$] & [pc] & [pc] & [$''$] &
[pc] & [pc] \\
\hline
05358+3543   & 2 & 0.02 & &    & & &  3  & 0.03 &  \\
18089$-$1732 & 0.5& 0.03 & 0.01 & 0.5 & 0.03 & 0.01 & 10 & 0.63 &  0.18\\
18151$-$1208 & 0.5& 0.01 & &    & & &    & &  \\
18182$-$1433 & 1 & 0.06 & 0.02 &    & & & 12 & 0.67 & 0.26  \\     
18264$-$1152 & 1 & 0.06 & 0.02 &    & & & 7  & 0.67 & 0.26  \\
18290$-$0924 & 2 & 0.1 & 0.01 & 20 & 1.02 & 0.51 & 1  & 0.05 & 0.03 \\
18306$-$0835 & 0.5& 0.03 & 0.01 & 12 & 0.62 & 0.29 & 12 & 0.62 & 0.29 \\
18308$-$0841 & 4 & 0.2 & 0.1 &    & & & 7  & 0.36 & 0.17 \\
18372$-$0541 & 1 & 0.07 & 0.01 &  1 & 0.07 & 0.01 & 1  & 0.07 & 0.01 \\
18385$-$0512 & 1 & 0.06 & 0.01 &  3 & 0.19 & 0.03 & 1  & 0.06 & 0.01 \\
18488+0000   & 4 & 0.17 & 0.11 & 20 & 0.89 & 0.54 & 3  & 0.13 & 0.08 \\
18553+0414   & 1 & 0.06 & 0.003 & 18 &1.13 & 0.05 & 3  &0.19 & 0.01 \\
18566+0408   & 0.5& 0.02 & &    & & & 1  & 0.02 &  \\
19012+0536   & 0.5& 0.02 & 0.01 &    & & & 7  & 0.29 & 0.16 \\
19035+0641   & 2 & 0.02 & & 7  &0.08  & & 1  & 0.01 &  \\
19217+1651   & 1 & 0.05 & &6/1 & 0.3/0.05 & & 2  &0.1 &  \\
19410+2336   & 5 & 0.16 & 0.05 & 1  & 0.03 & 0.01 & 5  & 0.16 & 0.05  \\
20126+4104   & 2 & 0.16 & &    & & & 2  & 0.16 &  \\
20293+3952   & 2 & 0.19 & 0.13 & 18 & 0.18 & 0.11 & 22 & 0.21 & 0.14 \\
23033+5951   & 1 & 0.02 & & 1.5& 0.03 & & 4  & 0.07 &  \\
23139+5939   & 1 & 0.02 & & 1  & 0.02 & & 0.5 & 0.01 &  \\
23151+5912   &1.5& 0.04 & &    & & & 7  & 0.19 &  \\
\hline
\end{tabular}
\caption{Separations of the H$_2$O masers from mm, cm and mid-infrared 
sources. Far and near values correspond to the far and near distance,
respectively. For sources without near values, the distance ambiguity
is solved \citep{sridha}. The {\bf uncertainties} are $\sim 5''$, scaling
with the distances for the linear separations. \label{h2o_sep}}
\end{table*}

\begin{table}
\begin{tabular}{lrrr}
association & number & $\%$ & mean\\ & & & [pc] \\
\hline 
CH$_3$OH/H$_2$O & 10   &  56  &  0.02 $\pm 0.03$ \\
CH$_3$OH/mm     & 18   &  100 &  0.03 $\pm 0.10$ \\ 
CH$_3$OH/cm     & 3    &  17  &  0.19 $\pm 0.03$ \\
CH$_3$OH/MIR    & 11   &  61  &  0.13 $\pm 0.10$ \\
H$_2$O/mm       & 22   &  100 &  0.03 $\pm 0.10$ \\
H$_2$O/cm       & 6    &  27  &  0.13 $\pm 0.03$ \\
H$_2$O/MIR      & 13   &  59  &  0.09 $\pm 0.10$ \\
\hline
\end{tabular}
\caption{Frequency of associated features within the spatial resolution 
as given in \S \ref{overall}. The third column gives the percentage of
associations with respect to the total number of sources where
CH$_3$OH or H$_2$O masers were detected. The last column presents the
mean separations and their uncertainties.}
\label{association}
\end{table}

\subsection{Associations and (non-)correlations}

Regarding the sources south of declination 20$^{\circ}$, which were in
the observable range of both telescopes (the ATCA and the VLA), and
including 19410+2336 and 20126+4104 (EVN and VLA data from
\citealt{minier 2000,minier 2001}), in 11 sources both maser types are
detected, in 7 just the CH$_3$OH masers, and in 6 only the H$_2$O
masers. Thus, $\sim 38\%$ of the CH$_3$OH maser sources do not show
H$_2$O maser emission, and $\sim 35\%$ of the H$_2$O maser sources do
not emit in the CH$_3$OH maser line. This shows that both maser types
could be powered by similar sources, but that finding one does not
necessarily imply finding the other as well. It rather confirms
significant differences in the emission process as already expected by
the different velocity ranges (e.g., \citealt{garay 1999},
\citealt{sridha}). As we are dealing with low-number statistics, we
estimate the error by $1/\sqrt{N}$ with N being the number of sources
we are investigating. Thus, the approximate statistical errors of our
analysis are in the $20\%$ range (relative to the percentages of
associations and (non-)correlations.) In sources with both maser
types, the masers seem to be always associated within the accuracy of
$1.5''$, except for $18151-1208$, the only field of view where both
types are detected but in two separate mm cores, $80''$ apart. In
contrast, in many sources just one of both maser types is present.

Both maser species are always found close to a mm core, traced by the
dust continuum peak, with an average projected separation of 0.03~pc
(Table \ref{association}). Considering the mean spatial separations
within star-forming cluster (see \ref{database}), it is possible that
both masers are amplified by the same powering source. Interestingly,
6 out of 22 H$_2$O masers ($27\pm 5\%$) are, within the observing
accuracy, also associated with cm emission, while only 3 out of 18
CH$_3$OH masers ($17\pm 5\%$) are nearby the cm peaks. The
statistical difference between both maser--cm associations is only
weak. It is quite possible that some of the maser/cm associations are
even chance alignments due to projection effects, but we cannot
quantify that more precisely. We note that we do not find any
source with CH$_3$OH and cm emission without an H$_2$O maser
nearby, whereas in three sources H$_2$O and cm peaks are coincident
within our accuracy, but no observed CH$_3$OH maser is found in the
same region. An interesting example, which illustrates the different
maser properties, is 19217+1651: some H$_2$O and CH$_3$OH features are
associated with the massive mm core, whereas another H$_2$O feature is
found near the cm source approximately $6''$ apart (Figure
\ref{maser1}). As the mm data are from interferometric Plateau de
Bure observations (accuracy $1''$), the separation of the cm and mm
peaks is real.

The mid-infrared sources stand somewhere between all the other
features. In some cases, they seem to be associated with the mm
sources and in other objects with the cm sources. The same is observed
for their spatial associations with both maser species, in some
sources their positions coincide within the observing accuracy, in
others they do not. A very peculiar example is 18488+0000, where an
archetypical cometary ultracompact H{\sc ii} region is at the edge of
the mm core, but most surprising, the mid-infrared source as well as
the H$_2$O maser feature do not coincide with any of them, but are
located at a small secondary mm peak that looks insignificant from the
mm point of view (but is still massive, $\sim 400$~M$_{\odot}$ at the
near kinematic distance of $\sim 5$~kpc, \citealt{beuther 2001b}).

To get a quantitative understanding of the spatial associations,
we tried to find correlations between the linear separations of the
different observed tracers and other quantities of our sample (the
luminosities, core masses, H$_2$ column densities and densities, as
taken from \citealt{sridha} and \citealt{beuther 2001a}). We did not
find any correlation among all those different quantities.

\subsection{Internal maser structure}
\label{internal}

To investigate whether the maser features show linear or arc-like
structures indicative of specific emission regions (disks, outflows,
or expanding shock waves, e.g., \citealt{walsh 1998}, \citealt{garay
1999}, \citealt{norris 2000}, \citealt{kylafis 1999}, and
\citealt{minier 2000}), we focused on the 6 sources for which we have
at least 3 spatially separate maser features in one or both
species. Outflows are ubiquitous features in massive star-forming
regions \citep{sridha,richer 2000,churchwell 2000}, therefore we
additionally present the bipolar outflow directions found by
\citet{beuther 2001b}. Figure \ref{maser_zoom} shows the results.

Morphologically, one can find rings, arcs or linear structures
in all cases, even between different maser species. But none of them
has a coherent velocity structure and therefore cannot easily be
interpreted in terms of disks, outflows or expanding shock
waves. There is also no obvious correlation with the outflow
direction. With only one set of observations, it is very difficult to
interpret these morphological and kinematic signatures, but it is well
possible that follow-up observations, and thus proper motion studies,
might be capable of setting better constraints on the regions where the
maser emission is produced. Such studies were successful in a
number of cases, e.g., \citet{alcolea 1992} and \citet{minier
2000b}.

\begin{figure}
\caption{Zoom into some maser distributions. The triangles show 
H$_2$O masers and the squares CH$_3$OH masers. The numbers mark the
radial velocities at which each feature emits, and the arrows
indicate the overall directions of the molecular outflows (solid: blue
wing, dashed: red wing) as found by \citet{beuther 2001b}. The axes
show the offsets in arcseconds from the IRAS position. 
\label{maser_zoom}}
\end{figure}


\section{Discussion}  

The observations presented in section \ref{overall} confirm that both
maser species~-- CH$_3$OH and H$_2$O~-- are signposts of very early
stages of massive star formation. While $100\%$ of the maser
detections are clearly associated with massive molecular cores traced
by mm dust continuum emission, only about $20\%$ are also associated
with cm continuum emission that is indicative of recently ignited, and
thus more evolved stars. These observations confirm results obtained
for each maser species separately over the last few years, that
CH$_3$OH and H$_2$O masers are signposts of very early stages of
massive star formation, and that both seem to cease emission soon
after the ignition of a central object (e.g., \citealt{tofani 1995},
\citealt{codella 1996,codella 1997}, \citealt{walsh 1998}, 
\citealt{minier 2001}). Here we show for the first time the similarities
and differences of both maser types within a consistent study of a
homogenous sample.

In spite of the small separations around 0.03~pc within many sources,
there are clear differences between both maser types. In addition to
their large velocity differences \citep{sridha}, in approximately
$30\%$ of the sources just one maser species is found. We do find a
few sources with nearby H$_2$O maser and cm emission, but not showing
CH$_3$OH maser emission. In contrast, every source, where CH$_3$OH
maser emission is associated with a cm source, has an H$_2$O maser
nearby. This difference suggests tentatively that CH$_3$OH maser
emission might be more sensitive to the ignited central object, and
may cease emitting in somewhat earlier evolutionary stages than H$_2$O
maser emission. However, it has to be noted that counterexamples
exist, e.g, the W3(OH) region, where the CH$_3$OH masers are
associated with the ultracompact H{\sc ii} region, and the H$_2$O
masers are approximately $6''$ to the east associated with the hot
core W3(H$_2$O). For a summary of the results in that region see
\citet{menten 1996}.

CH$_3$OH maser models propose that the maser emission is produced
by radiative pumping in moderately hot regions ($\sim 150$~K) with
CH$_3$OH column densities $>2\times 10^{15}$~cm$^{-2}$ and hydrogen
number densities $<10^8$~cm$^{-3}$ \citep{hartquist
1995,sobolev 1997,cragg 2001}. The data presented here as well as
other studies (e.g., \citealt{walsh 1997,walsh 1998},
\citealt{phillips 1998}, \citealt{minier 2000,minier 2001}) indicate that
the ignition of a massive star, and thus the formation of an
ultracompact H{\sc ii} region, alters these conditions swiftly and
stops further CH$_3$OH maser emission very soon.

In contrast, H$_2$O maser models mostly explain the excitation by
collisional pumping with H$_2$ molecules in shocks associated with
outflows \citep{elitzur 1989}. Such collisional pumping can also occur
by accretion shocks in accretion disks \citep{garay 1999}. These
processes may not depend as strongly on the influence of the H{\sc ii}
region as the radiative pumping of the CH$_3$OH maser, and H$_2$O
maser emission might continue some time in more evolved stages of
evolution. But obviously, the statistics of the observations are not
sufficient for a stronger statement.

The percentage of associated maser and mid-infrared sources ($\sim
60\%$) lies in a similar regime as found recently by \citet{walsh
2001}. They propose that the mid-infrared objects, which are likely to
be embedded luminous protostellar objects, could be the pumping
sources of the maser sites.  This might be true in the associated
sources of our sample as well, but the percentage of associations
clearly shows that detections in the mid-infrared regime are not
necessary to produce the maser emission. It is possible that for high
obscuration the sources are not detectable in the mid-infrared -- only
the maser shows up.

Additionally, we do not find strong correlations between the spatial
separations of the different observed quantities and other
characteristic quantities of the sources, e.g., the luminosities, core
masses, column densities, and volume densities. Thus, it seems likely
that both maser species need a similar environment of dense and warm
gas, but that, due to the different excitation processes (radiative
pumping for CH$_3$OH and collisional pumping for H$_2$O), no real
correlation exists. Spatial associations of both maser species might
be just coincidences caused by insufficient spatial resolution and
projection effects.

We cannot determine accurately in any source of this sample whether
the maser emission is produced in disks, outflows or shock waves. This
stresses that kinematic interpretations of different maser features
are difficult and not as straightforward as sometimes supposed in the
past (e.g., \citealt{garay 1999}, \citealt{norris
1998}). CH$_3$OH and H$_2$O masers are excellent signposts of massive
star-forming regions, but the detailed interpretation of the different
kinematic and spatial features has to be treated with caution. There
are definitely cases where the kinematic information gives deep
insights in special sources, e.g., the H$_2$O maser in W3(H$_2$O)
\citep{alcolea 1992}, or some of the linear structures found in
CH$_3$OH by, e.g, \citet{walsh 1998},
\citet{minier 2000}, \citet{norris 1998}, and \citet{phillips 1998},
but as stressed by \citet{minier 2000}, proper motions and VLBI
observations are needed to derive conclusive answers for such
sources. We believe that the approach of kinematic interpretation of
different maser features works only in a limited number of sources
with a favorable geometry with respect to the observer, and then
especially when proper motion observations are available. With this
regard, the data presented in this paper are a well suited basis for
follow-up proper motion studies of both maser species. Except of this,
for most sources, the main and highly important outcome of maser
observations in massive star-forming regions is their characteristics
as signposts for high-mass star formation.


\begin{acknowledgements} 
We like to thank the referee Dr. G. Fuller for detailed comments on
the first draft, which improved the quality of the paper
significantly. H. Beuther got support by the {\it Deutsche
Forschungsgemeinschaft, DFG} project number SPP 471.
\end{acknowledgements}


\end{document}